\documentclass[onecolumn,nofootinbib,showpacs,aps,10pt]{revtex4}

\usepackage{amssymb}
\usepackage{amsmath}
\usepackage[dvips]{color}
\usepackage{bm}
\usepackage{graphics,graphicx}

\newcommand{\be}{\begin{equation}}
\newcommand{\ee}{\end{equation}}
\newcommand{\bea}{\begin{eqnarray}}
\newcommand{\eea}{\end{eqnarray}}

\newcommand{\sumint}{\sum\!\!\!\!\!\!\!\!\int}

\begin{document}

\title{Magnetic Field Effect on Charged Scalar Pair Creation at Finite Temperature}


\author{Gabriella Piccinelli$^*$, Angel S\' anchez$^\dagger$}{\affiliation{$^*$Centro Tecnol\'ogico, FES Arag\' on, Universidad Nacional Aut\' onoma de M\' exico, Avenida Rancho Seco S/N, Col. Impulsora, Nezahualc\' oyotl, Estado de M\' exico 57130, M\' exico.\\
$^\dagger$Facultad de Ciencias, Universidad Nacional Aut\' onoma de M\' exico, Apartado Postal 50-542, Ciudad de M\'exico 04510, M\'exico.}

\begin{abstract}
In this work we explore the effects of a weak magnetic field and a thermal bath on the decay process of a neutral scalar boson into two charged scalar bosons. Our findings indicate that magnetic field inhibits while temperature enhances the pair production. 

The employed formalism allows us to isolate the contribution of magnetic fields in vacuum, leading to a separate analysis of the effects of different ingredients. This is essential since the analytical computation of the decay width necessarily needs of some approximation and the results that can be found in the literature are not always coincident. We perform the calculation in vacuum by two different weak field approximations. The particle pair production in vacuum was found to coincide with finite temperature behavior, which is opposite to results obtained by other authors in scenarios that involve neutral particles decaying into a pair of charged fermions. Among other differences between these scenarios, we traced that the analytical structure of the self-energy imposed by the spin of particles involved in the process is determinant in the behavior of the decay rate with the magnetic field.  

\end{abstract}


\maketitle

\section{Introduction}

Nowadays, it is widely accepted that particle properties are modified under extreme conditions of high densities and temperatures and under the influence of magnetic fields making, high energy physics experiments, astrophysical compact objects and early universe events  excellent laboratories for exploring the effect of these external agents.

In particular, the effect of magnetic fields on a particle decay process have been extensively studied,  in many contexts: high intensity laser experiments~\cite{lasers}, relativistic heavy ion collisions (QCD)~\cite{Chernodub,Kawaguchi,Bandyopadhyay}, compact objects~\cite{Raffelt,compactobjects} and early universe events~\cite{BPS}. An interesting review can be found in~\cite{Borisov}. 

In the cases of study presented in the literature there are some basic differences. On one hand, the decay products can be fermions or bosons and, in the last case, scalars or vectors. On another hand, two limits are typically considered: strong magnetic fields, in which case only the lowest Landau level (LLL) is taken into account, and weak magnetic fields, that allow to perform some kind of expansion series in $B$ (with $B$ the magnetic field) and keep only the lowest terms.~Some situations are treated at finite temperature and, in other cases, temperature is neglected. Finally, the methods followed to calculate the decay rate and the approximations necessarily accomplished to go through the calculation vary from a work to another. Although these basic differences have to be taken into account, it is nonetheless remarkable that many of the results found in the literature do not coincide. We will try here to shed some light on the basic physical ingredients that can lead to a different behavior and on the possible mismatches introduced by different approximations (see also \cite{FelixKarbstein}). 

There are different approaches that take into account the influence of external magnetic fields on the particle creation process. A first effect is the magnetic correction to the mass of the decay products through the real part of the self-energy, thus shifting the decay threshold for particle creation. On another hand, the effect can be analyzed on the decay rate of the progenitor particle through the imaginary part of the self-energy or via a Bogoliubov transformation.

We have found that published results arrive to different conclusions: the decay process is enhanced by the magnetic field in some cases \cite{TsaiErber,Urrutia,Kuznetsov1,Mikheev,Erdas,Bhattacharya,Satunin,Bandyopadhyay}, inhibited in others \cite{Sogut,Kawaguchi} and can even present a mixed behavior for different energies \cite{Chistyakov,Ghosh} .

 As it was pointed out in Ref.~\cite{Kawaguchi}, an important difference may be the spin of the decaying particle. They found that magnetic fields have inverse effects on the decay rate of the $\rho$ meson for different polarization modes. When the $s_z=0$ mode decays, as the magnetic field increases, the decay width starts to develop, reaches a peak and then decreases, while for the $s_z \pm 1$ channel, the decay rates monotonically increase. Notice that they work at the LLL although their magnetic fields are not particularly high and get some inconsistencies due to the Landau levels truncation. Conversely,  in Ref.~\cite{Sogut} spin does not seem to play any role in the decay process. In their work, the pair creation of spin-1/2 particles in presence of electromagnetic fields is studied numerically via the Bogoliubov transformation method, finding that the particle creation process in a constant magnetic field is inhibited.  The two-component formalism for the Dirac equation they used allowed them to extend this conclusion to the decay of  $s=0$ particles. Still a different behavior is obtained in \cite{Bandyopadhyay} where, to leading order in $B$, the $\rho^0$ meson decay width is found to increase with the magnetic field.

In order to explore the effect of external magnetic fields on the decay process and its possible relation with spin, 
in this work we shall study a heavy scalar boson decay into two charged scalar particles, in vacuum as well as at finite temperature, with different approaches. In particular, we are  interested in weak field limit, in such a way that a direct application of our study could be the inflaton decay process in a warm inflation scenario \cite{Berera,Bastero-Gil,HM},  considering that cosmic magnetic fields observed at all scales in the universe~\cite{cosmicB,IntergalacticB}  could be primordial~\cite{SavvidyEnqv-Olesen,TurnerWidrow,PlanckB}.

The outline of this work is as follows: in Sec.\ref{sec2} we introduce the magnetic field and temperature effects on charged scalar propagators and, within a theory of three scalar bosons interaction, we get the heavy boson self-energy in the weak field limit; in Sec.\ref{sec3} the imaginary part of the self-energy, presented in the previous section, is calculated and its relation with the charged scalar pair creation is established.  We present the magnetic field effect on the scalar pair creation in vacuum with two different approaches in Sec.\ref{sec4}, discussing the physical differences between these two approaches and with respect to scenarios analyzed by other authors. Finally, Sec.\ref{sec5} contains our conclusions.

\section{Propagators and self-energies at finite temperature in a uniform weak magnetic field}\label{sec2}

Let us consider a model in which a neutral scalar boson $\Phi$ interacts with two charged scalar bosons $\phi^-\phi^+$.  
The Lagrangian that accounts for this process could have the form
\begin{equation}
   \mathcal{L}=g\Phi\phi^*\phi.
\end{equation}
This interaction term between the heavy boson and the light ones gives rise to the Feynman diagram shown in Fig.\ref{fig1}, whose analytical expression is given by
\bea
  i \Pi(p)=(ig)^2\int\frac{d^4k}{(2\pi)^4}D_{B}(p-k)D_B(k).
\label{pivacdef}
\eea
\begin{figure}[h!]
\begin{center}
  \includegraphics[width=0.3\textwidth]{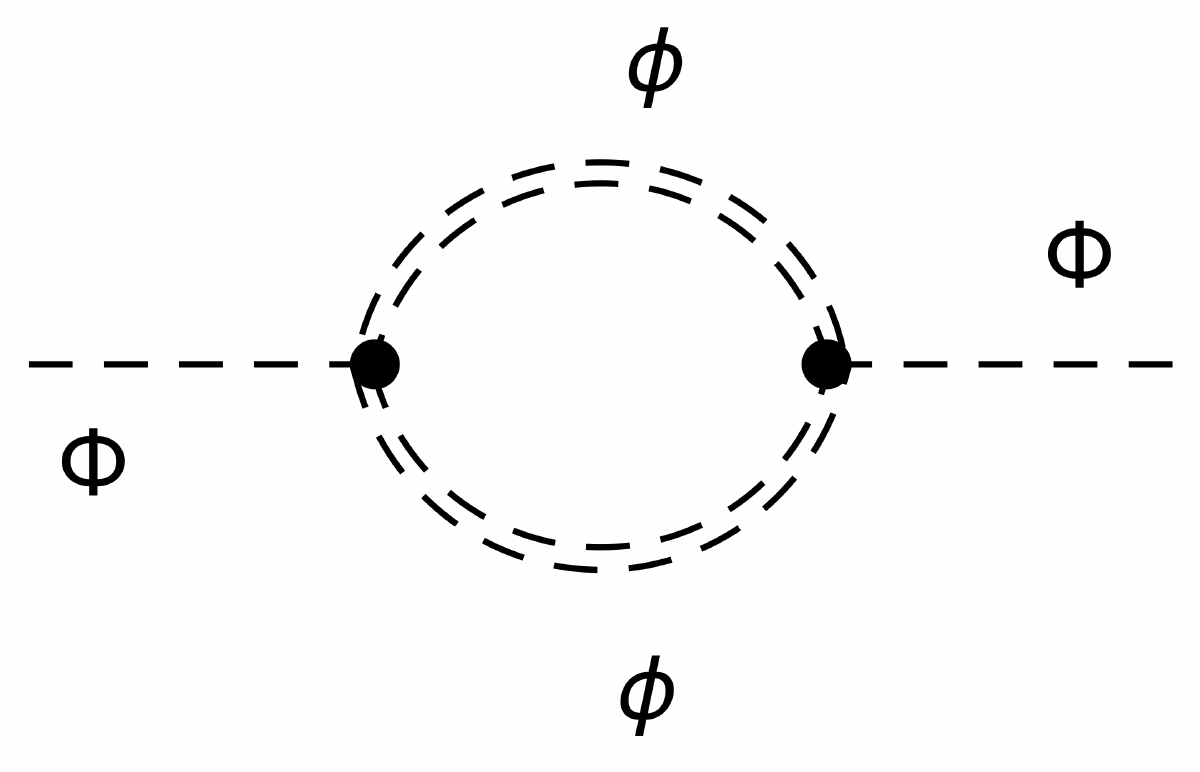} 
    \end{center}
    \caption{Leading-order contribution to the neutral scalar $\Phi$  boson self-energy in a magnetic field background. $\phi's$ are light charged scalar bosons and  the double dashed lines represent their propagators dressed with the magnetic field.}
\label{fig1}
\end{figure}
The effect of an external magnetic field is incorporated by  using
Schwinger's proper-time method~\cite{Schwinger}, where the momentum
dependent propagators for charged scalars coupled to the
external field take the form 
\begin{eqnarray}
\label{scalpropmom}
D_{B} (k) &=& \int_0^\infty \frac{ds}{\cos{eBs}}  \exp\left\{ i s \left(k_{||}^2-k_{\bot}^2 \frac{\tan{eBs}}{eBs}-m^2 +i \epsilon\right)\right\},
\end{eqnarray}
where $e$ and $m$ denote the charge and mass of the scalar field $\phi$, respectively, and $B$ is the external magnetic field.
Since we are considering an external uniform magnetic
field along $z$-direction, the notation we are adopting is $k_{||}^2\equiv k_0^2-k_3^2$ and $k_\perp^2\equiv k_1^2+k_2^2$.

Let us restrict ourself to the hierarchy of scales  $eB\ll m^2\ll T^2$ which could be relevant in some of the contexts discussed in the introduction. In this framework, the summation over Landau levels can be performed in such a way that  a  weak field expansion in Eq.~(\ref{scalpropmom}) can be achieved~\cite{Chyi2000,Ayala2005}.  This allows us to write the scalar  propagator as power series in $eB$ which, up to order $(eB)^2$, reads 
\begin{eqnarray}
\label{scalpropweak}
D_{B}(k)&\simeq&\frac{i}{k^2-m^2}-(eB)^2\left(\frac{i}{(k^2-m^2)^3}
+ \frac{2ik_\perp^2}{(k^2-m^2)^4}\right).
\end{eqnarray}
Thermal effects are incorporated, within the framework of the imaginary time formalism,  by making the replacements
\bea
     D(k) \rightarrow \Delta(K)\equiv\frac{1}{K^2+m^2}=\frac{1}{\omega_n^2+\omega^2} \hspace{1cm}\mbox{and} \hspace{1cm} 
     \int\frac{d^4k}{(2\pi)^4} \rightarrow \sumint\frac{d^4K}{(2\pi)^4}\equiv\frac{1}{\beta}\sum_{n=-\infty}^\infty \int\frac{d^3k}{(2\pi)^3},
\eea
where $\omega_n=2\pi n/ \beta$ are the Matsubara frequencies, $T=1/\beta$ the temperature and $\omega^2=k^2+m^2$.

Thus,  at finite temperature and in the weak field approximation, the self-energy of the scalar field $\Phi$ becomes 
\bea
&&\hspace{-0.4cm}\Pi_{T,B}(P)=-g^2\sumint\frac{d^3k}{(2\pi)^3}\Bigg\{\Delta(P-K)\Delta(K)
\nonumber \\
&&-(eB)^2\Delta(P-K)\Delta(K)\bigg(\Delta^2(K)+\Delta^2(P-K)
\nonumber \\
&&-2k_\perp^2\Delta^3(K)-2(k-p)_\perp^2\Delta^3(P-K)\bigg)\Bigg\}, 
\label{selfenergyTB}
\eea
where we adopted the notation: 4-momenta are written in upper case and 3-momenta in lower
case. 

To show the main steps used in the calculation of Eq.(\ref{selfenergyTB}), let us calculate explicitly the first term in the self-energy
\bea
      \Pi_0(P)&=&-g^2\sumint\frac{d^4K}{(2\pi)^4} \frac{1}{4\omega_1\omega_2}\sum_{r,s=\pm1}\frac{rs}{i\omega_m+r\omega_1+s\omega_2}\left(\frac{1}{-i\omega_n+i\omega_m+r\omega_1}+\frac{1}{i\omega_n+s\omega_2}\right),
\label{pi0sum}
\eea
where $\omega_1^2\equiv(p-k)^2+m_1^2$ and $\omega_2^2\equiv k^2+m_2^2$ and we have used partial fraction decomposition and two different masses for computational advantages.

By using the identity~\cite{Kapusta}
\bea
\frac{1}{\beta}\sum_{n=-\infty}^\infty \frac{1}{i\omega_n+s \omega}=s \left(n(\omega)+\frac{1}{2}\right),
\label{matsubarasum}
\eea
with $n(\omega)$ the Bose-Einstein distribution, the sum over Matsubara frequencies can easily be done in Eq.(\ref{pi0sum}), and we get
\bea
      \Pi_0(P)&=&-g^2\int\frac{d^3k}{(2\pi)^3} \frac{1}{4\omega_1\omega_2}\sum_{r,s=\pm1}\frac{1}{i\omega_m+r\omega_1+s\omega_2}
                    \left(\frac{r+s}{2}+s\ n(\omega_1)+r\ n(\omega_2)\right),
\label{pifullzeroB}                    
\eea
where the periodical conditions have been accounted for. 

The rest of terms in Eq.(\ref{selfenergyTB}) are computed in a similar fashion by noticing that the expansion performed in power series of $eB$ has  products of propagators elevated to different powers, which in turn translates into having derivatives of the boson distribution function~\cite{warmus}, through 
\bea
      \Delta^{n+1}(K)&\equiv&\frac{1}{(K^2+m_i^2)^{n+1}}
       \nonumber \\
       &=&\frac{(-1)^n}{n!}\frac{\partial^n}{\partial (m_i^2)^n} \Delta(K).
\label{trick}
\eea
According to the imaginary time formalism, once the sum over the Matsubara frequencies is done, one is allowed to make an analytical continuation of the external momentum: $i\omega_m\rightarrow p_0+i\epsilon$~\cite{LeBellac}. This analytical continuation allows us to study the imaginary part of the self-energy, and this will be done in the next section.

\section{Magnetic Field effect on Scalar Decay width at finite temperature}\label{sec3}

\begin{figure}[h!]
\begin{center}
  \includegraphics[width=0.3\textwidth]{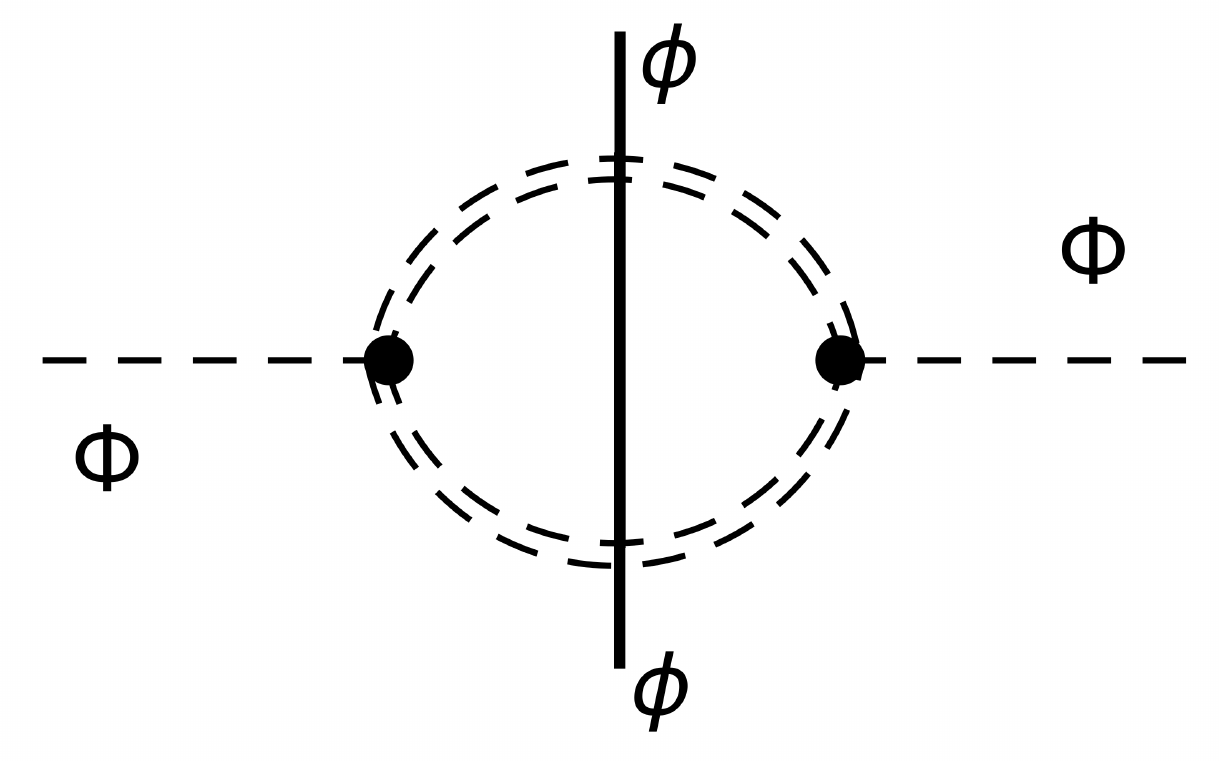} 
    \end{center}
    \caption{Feynman representation for the imaginary scalar self-energy (indicated by the vertical line).}
\label{fig2}
\end{figure}

The imaginary part of Eq.(\ref{pifullzeroB}), whose diagrammatic representation is shown in Fig.\ref{fig2}, can easily be calculated through
\bea
      \Im\Pi_0(p)&=&\frac{\Pi(p_0+i\epsilon)-\Pi(p_0-i\epsilon)}{2i},
\eea
obtaining
\bea
     \Im\Pi_0(p)&=&g^2\pi\int\frac{d^3k}{(2\pi)^3} \frac{1}{4\omega_1\omega_2}\sum_{r,s=\pm1}\delta(p_0+r\omega_1+s\omega_2)
                    \left(\frac{r+s}{2}+s\ n(\omega_1)+r\ n(\omega_2)\right),
\label{impivac0}
\eea
where we have employed the Dirac delta representation (see e.g.,~\cite{LeBellac})
\bea
     \delta(x)= \frac{1}{\pi}\lim_{\epsilon\rightarrow0}\frac{\epsilon}{(p_0+r\omega_1+s\omega_2)^2+\epsilon^2}.
\eea

Note that Eq.(\ref{impivac0}) contains several processes related with the interaction of the heavy scalar with the thermal bath, nevertheless, since we are interested in the decay width of the scalar boson, we shall focus only on the term $r=s=-1$~\cite{weldon}, which, in a symmetric form in  $\omega_1$ and $\omega_2$, is explicitly given by
\bea
     \Im\Pi_0(p)=-g^2\pi\int\frac{d^3k}{(2\pi)^3} \frac{1}{4\omega_1\omega_2}
                          \left[ \left(\frac{1}{2}+ n(\omega_1)\right)\delta(p_0-\omega_1-\omega_2)
                          +\left(\frac{1}{2}+ n(\omega_2)\right)\delta(p_0-\omega_1-\omega_2)\right].
\label{impi0decay1}                   
\eea
This decomposition will be useful when  studying the magnetic effect on the self-energy. 

Performing the angular integration on each term in Eq.(\ref{impi0decay1}), we obtain
\bea
     \Im\Pi_0(p)&=&-\frac{g^2\pi}{(4\pi)^2 |\vec{p}|}
                            \left[
                                   \int_{\omega_1^-}^{\omega_1^+} d\omega_1 \left(\frac{1}{2}+ n(\omega_1)\right) +
                                   \int_{\omega_2^-}^{\omega_2^+} d\omega_2 \left(\frac{1}{2}+ n(\omega_2)\right) 
                           \right]\theta(p\cdot p-(m_2+m_1)^2)  
\nonumber \\
                        &=&-\frac{g^2}{16\pi  |\vec{p}|}
                          \sum_{\stackrel{i=1,2}{\sigma=\pm1}}
 			  \sigma\left[
  			 \frac{\omega_i^\sigma}{2}+\frac{1}{\beta}\ln\left(1-e^{-\beta \omega_i^\sigma}\right)
			 \right]\theta(p\cdot p-(m_2+m_1)^2),
\label{vacuumsymm}			 
\eea
where 
\bea
     \omega_i^\pm\equiv  \frac{p_0[p\cdot p+(-1)^i(m_2^2-m_1^2)] }{2(p\cdot p)}
         \pm|\vec{p}|\frac{\sqrt{[p\cdot p-(m_1-m_2)^2][p\cdot p-(m_1+m_2)^2]}}{2(p\cdot p)} ,
\eea  
with $p\cdot p=p_0^2-\vec{p}^{\ 2}$. This result is the symmetric form of the imaginary part of  a boson self-energy obtained in Ref.~\cite{Bastero-Gil}. 

Repeating the analysis done in $\Im\Pi_0(p)$ to the remaining terms in Eq.(\ref{selfenergyTB}) together with Eq.(\ref{trick}), the
imaginary part of the self-energy  that accounts for magnetic field and temperature effects reads
\begin{eqnarray}
  &&\hspace{-0.7cm} \Im\Pi_{T,B}(p)=-\frac{g^2}{16 \pi|\vec{p}|}\sum_{\stackrel{i=1,2}{\sigma=\pm1}}
   \sigma\Bigg\{
   \frac{\omega_i^\sigma}{2}+\frac{1}{\beta}\ln\left(1-e^{-\beta \omega_i^\sigma}\right)
\nonumber \\
&&+\frac{(eB)^2}{12}\Bigg[-\frac{1}{12{\omega_i^\sigma}^3}+\frac{m_i^2(1+2n(\omega_i^\sigma))}{2{\omega_i^\sigma}^5}+\frac{n(\omega_i^\sigma)}{2{\omega_i^\sigma}^3}
\nonumber \\
&&-\frac{1}{2{\omega_i^\sigma}^2}\left.\frac{d n(\omega_i)}{d\omega_i}\right|_{\omega_i^\sigma}
+m_i^2\left.\frac{d}{d\omega_i}\left(\frac{n(\omega_i)}{\omega_i^4}\right)\right|_{\omega_i^\sigma}
\nonumber \\
&&-\frac{1}{3}\left.\frac{d^2}{d\omega_i^2}\left(\frac{(\omega_i^2-m_i^2)n(\omega_i)}{\omega_i^3}\right)\right|_{\omega_i^\sigma}\Bigg]\Bigg\}
\theta(p\cdot p-(m_1+m_2)^2).
\label{IMpifullTB}
\end{eqnarray}
Note that in this result there is not an explicit dependence on $p_\perp$ since we have approximated  $k_\perp^2$ as $\frac{2}{3}k^2$  in integrals that involve transverse momentum. This approximation is valid in the weak magnetic field limit. 

To get the imaginary part of the process described in Fig.(\ref{fig2}), now we set $m_1=m_2=m$ in Eq.(\ref{IMpifullTB}), obtaining
\bea
    &&\hspace{-0.7cm}\Im\Pi_{T,B}(p)=-\frac{g^2}{16 \pi|\vec{p}|}\sum_{\sigma=\pm1}
   \sigma\Bigg\{
   \omega^\sigma+\frac{2}{\beta}\ln\left(1-e^{-\beta \omega^\sigma}\right)
\nonumber \\
&&+\frac{(eB)^2}{12}\Bigg[-\frac{1}{6{\omega^\sigma}^3}+\frac{m^2(1+2n(\omega^\sigma))}{{\omega^\sigma}^5}+\frac{n(\omega^\sigma)}{{\omega^\sigma}^3}
\nonumber \\
&&-\frac{1}{{\omega^\sigma}^2}\left.\frac{d n(\omega)}{d\omega}\right|_{\omega^\sigma}
+2m^2\left.\frac{d}{d\omega}\left(\frac{n(\omega)}{\omega^4}\right)\right|_{\omega^\sigma}
\nonumber \\
&&-\frac{2}{3}\left.\frac{d^2}{d\omega^2}\left(\frac{(\omega^2-m^2)n(\omega)}{\omega^3}\right)\right|_{\omega^\sigma}\Bigg]\Bigg\}
\theta(p\cdot p-4m^2),    
\label{decayfullTB}
\eea
where 
\bea
\omega^\sigma=\frac{p_0}{2}+\sigma \frac{|\vec{p}|}{2}\sqrt{1-\frac{4m^2}{p\cdot p}}.
\eea
As we are interested in the effect of a thermal bath and the magnetic field effect on the decay process of the heavy particle, we can now easily compute the decay width through the optical theorem~\cite{peskin} 
\bea
      \Gamma_{T,B}=-\frac{\Im \Pi_{T,B}(p)}{\sqrt{\vec{p}^{\ 2}+M^2}},
\label{decaydef}
\eea
with $M$ the heavy decaying boson mass. Note that at finite temperature the effective decay process $\Gamma$ is obtained in the same way and accounts for two contributions  $\Gamma = \Gamma_d-\Gamma_i$,
where $\Gamma_d$ is the decay rate and $\Gamma_i$ the inverse decay \cite{weldon}, which can be neglected when $T\ll M$. Additionally, observe that the magnetic field and temperature do not affect the particle production threshold (fixed by the argument of theta function in Eq.(\ref{decayfullTB})) since we have not considered corrections to the light particle masses.  In the case we would account for this corrections, the threshold would be shifted to higher or lower energies \cite{Chernodub, warmus}.

Since the imaginary time formalism contains the vacuum contribution, we can isolate it by setting $T=0$, getting in the rest frame of the decaying particle
\bea
    \Gamma_{0,B}= \frac{g^2}{16 \pi}\frac{1}{M}\sqrt{1-\frac{4m^2}{M^2}}\left[ 1+\frac{2}{3}\frac{(eB)^2}{M^4}\left(1-40\frac{m^2}{M^2}\right)\right]\theta(p\cdot p-4m^2).
\label{decayvacB}
\eea
Note that the first term corresponds to the vacuum result for the two body decay process shown in Eq.~(\ref{ImPivacuum}).

In Fig.(\ref{decayvac1}) we plot the decay width behavior given by Eq.(\ref{decayvacB}) as a function of the magnetic field in vacuum. In this figure and the following ones we shall ignore the factor  $g^2/16\pi$. Note that as the magnetic field increases the decay width becomes suppressed.
\begin{figure}[h!]
\begin{center}
  \includegraphics[width=9cm]{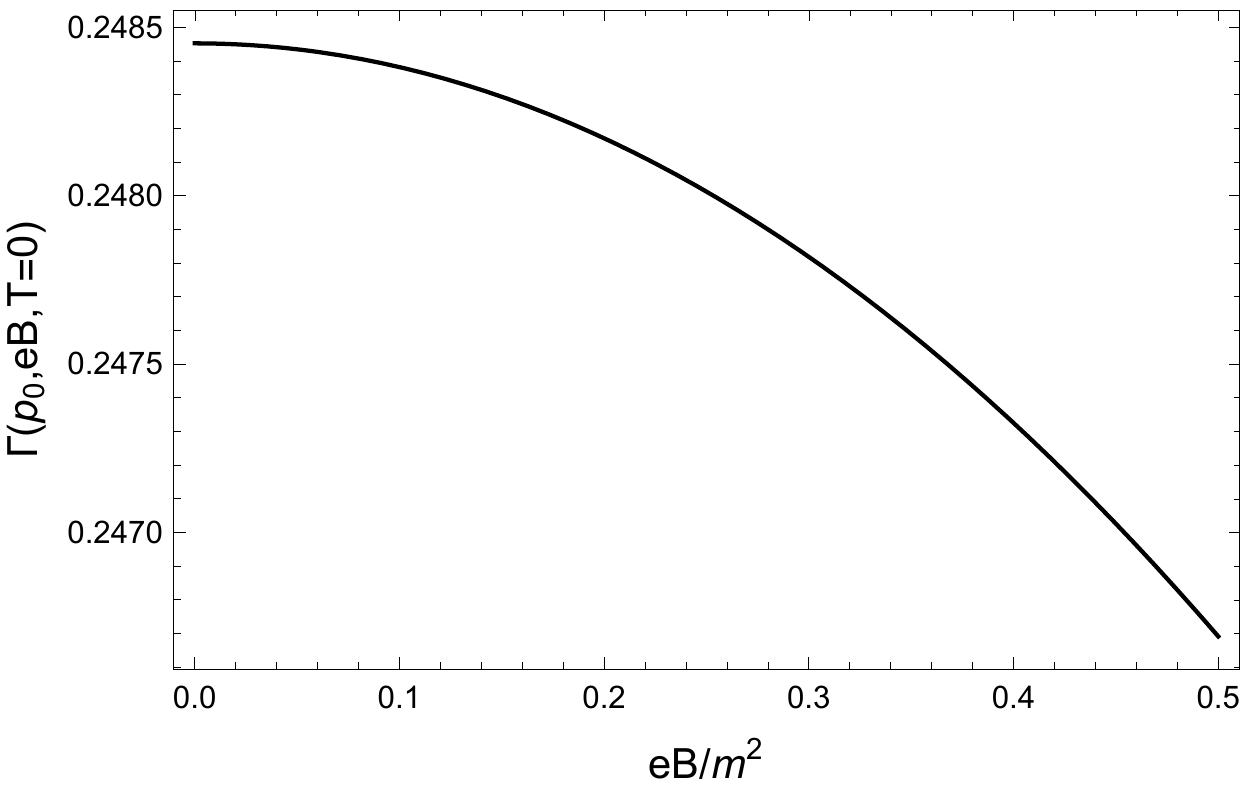} 
    \caption{({color online}) Magnetic field effect of heavy boson decay rate at $T=0$}
\label{decayvac1}
\end{center}
\end{figure}

\begin{figure}[h!]
\begin{center}
  \includegraphics[width=8.7cm]{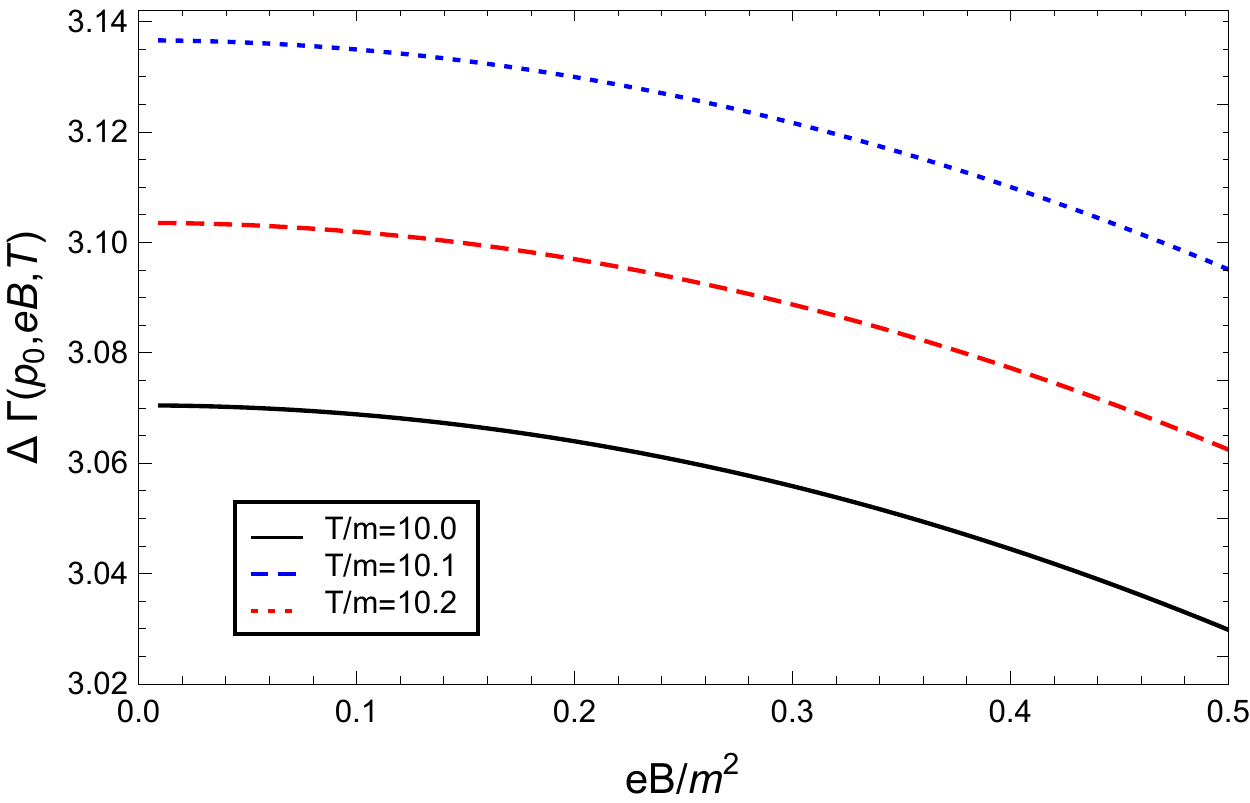} 
    \caption{({color online}) Magnetic field effect on heavy boson decay width for different temperatures.}
\label{deltadecayTB}
\end{center}
\end{figure}
In order to study the effect of a magnetized medium at finite temperature on the decaying process, we subtract the vacuum contribution given in Eq.(\ref{decayvacB}) from the full expression that can be obtained from Eq.(\ref{decayfullTB}): 
\bea
     \Delta \Gamma_{T,B} (p)=\Gamma_{T,B}(p)-\Gamma_{0,B}(p).
\eea
In particular, in  the decaying particle rest frame, this effect is shown in Fig.(\ref{deltadecayTB}). As in the vacuum case, the magnetic field  suppresses the scalar decay process. On the other hand, temperature enhances the process,  and this can be understood as follows: the available number of states of the heavy boson, which is proportional to $n(E)$, grows as a function of temperature, increasing the phase space for the process.

As it was pointed out in the introduction, the weak field expansion has been developed in different ways by some authors, obtaining results not always coincident. Since in this section we have dealt with the magnetic field and temperature, then  in the next one we explore only the effect of an external magnetic field on the decay process.

\section{Magnetic Field Effect on Scalar decay in vacuum}\label{sec4}

The scalar self-energy for the $\Phi$ boson in vacuum, Eq.~(\ref{pivacdef}), which accounts for a magnetic field background, has the form 
\bea
i \Pi_B(p)=-g^2\int \frac{d^4k}{(2\pi)^4}\int_0^\infty \frac{ds_1 ds_2}{\cos(eBs_1)\cos(eBs_2)}
          e^{-i(s_1+s_2)m^2} 
          e^{is_1[(p-k)_{||}^2-(p-k)_{\perp}^2\frac{\tan(eBs_1)}{eBs_1}]}
          e^{is_2[(k)_{||}^2-k_{\perp}^2\frac{\tan(eBs_2)}{eBs_2}]}.
\nonumber \\ 
\eea
Once the gaussian integration is carried out over the loop momentum $k$, we get
\bea
\Pi_B(p)&=&\frac{g^2}{16\pi^2}\int_0^\infty \frac{ds_1 ds_2}{\cos(eBs_1)\cos(eBs_2)}\frac{eB\ e^{-i(s_1+s_2)m^2 }}{(s_1+s_2)(\tan(eBs_1)+\tan(eBs_2))}
       e^{i\left(\frac{s_1s_2}{s_1+s_2}p_{||}^2-p_{\perp}^2\frac{\tan(eBs_1)\tan(eBs_2)}{\tan(eBs_1)+\tan(eBs_2)}\right)}.
\nonumber \\ 
\eea
Making the replacements on the proper time variables \cite{TsaiErber}
\bea
      s_1\rightarrow s\frac{1-v}{2} 
\hspace{1cm}\mbox{and}\hspace{1cm}
      s_2\rightarrow s\frac{1+v}{2},
      \nonumber 
\eea
the scalar self-energy becomes
\bea
     \Pi_B(p)=\frac{g^2}{32\pi^2}\int_0^\infty ds\int_{-1}^1dv \frac{eB}{\sin(eBs)}
          e^{-i s m^2} 
          e^{is\left[\frac{1}{4}(1-v^2)p_{||}^2-p_{\perp}^2\frac{\cos(eBsv)-\cos(eBs)}{2eBs\sin(eBs)}\right]}.    
\eea

Although the ultraviolet divergence in the self-energy does not affect the imaginary part, let us isolate it for simplicity. Following Ref.(\cite{Schwinger}), we perform an integration by parts over $v$,  obtaining
\bea
  &&\hspace{-0.2cm}  \Pi_B(p)=\frac{g^2}{16\pi^2}\int_0^\infty \frac{ds}{s}\ e^{-i s m^2}
      \nonumber \\
      &&+\
     \frac{ig^2 p^2}{32\pi^2}\int_{0}^1dv v^2 \int_0^\infty \frac{ds\ eBs}{\sin(eBs)}
           \left[1+\frac{p_\perp^2}{p^2}\left(1-\frac{\sin(eBsv)}{v\sin(eBs)}\right)\right]
          e^{is\left[\frac{1}{4}(1-v^2)p^2-m^2\right]}
          e^{isp_{\perp}^2\left[\frac{1}{4}(1-v^2)-\frac{\cos(eBsv)-\cos(eBs)}{2eBs\sin(eBs)}\right]}.
      \nonumber \\
\label{selfenergyfull}
\eea
Note that this result is valid for an arbitrary magnetic field strength. In what follows, we shall perform two different weak field approximations on Eq.(\ref{selfenergyfull}).

\subsection{Approximation {\it \`a la} Tsai and Erber}

In order to explore the weak field limit $eB\ll m^2$ on Eq.(\ref{selfenergyfull}) in this approximation, we take into account that the main contribution to the integral over $s$ comes from the region $eBs \ll 1$ \cite{TsaiErber,Urrutia}, then we carry out a Taylor expansion with care: the argument in the exponential that involves the magnetic field  is expanded up to $(eBs)^2$, however, since there are terms that contain a factor $p_\perp^2$ that can be large (known as crossed field approximation), then, the exponential itself cannot be expanded in powers of $eBs$.
This argument does not apply to the coefficient in front of the exponential since the transverse momentum is weighted by the total momentum, thus, the leading contribution is of the order $\mathcal{O}(1)$. Bearing this in mind, we get
\bea
   \Pi_B(p) = \frac{g^2}{16\pi^2}\int_0^\infty\frac{ds}{s}e^{-is m^2}
    + \frac{ig^2p^2}{32\pi^2}\int_0^1dvv^2\int_0^\infty ds e^{is[\frac{1}{4}(1-v^2)p^2-m^2]}
                 e^{-isp_\perp^2 \frac{1}{48}(1-v^2)^2(eBs)^2}.
\label{tsai1}
\eea
The imaginary part of the above equation reads 
\bea
\Im{\Pi_B(p)}
  &=&\frac{g^2p^2}{32\pi^2}\int^1_{\sqrt{1-\frac{4m^2}{p^2}}} dv \frac{\sqrt{m^2-\frac{1}{4}(1-v^2)p^2}}{(1-v^2)|p_\perp|eB}4v^2\int_0^\infty dy \cos\left(\frac{3}{2}\rho(y+\frac{1}{3}y^3)\right),
\eea
where we made the change of variable 
\bea
s= \frac{\sqrt{m^2-\frac{1}{4}(1-v^2)p^2}}{(1-v^2)|p_\perp|eB} y  
\eea
and introduced the notation
\bea
\nonumber \\
\rho\equiv\frac{4}{\lambda}\frac{\left(1-(1-v^2)\frac{p^2}{4m^2}\right)^{3/2}}{1-v^2} \ , \hspace{1cm}\mbox{with}\hspace{1cm}  \ 
\lambda \equiv \frac{3}{2}\frac{p_\perp}{m}\frac{eB}{m^2}\ ,
\eea
and the lower limit in the integration over $v$ ensures that the integrand in this region be a real number.

Identifying the integration over $y$ as the integral representation of Modified Bessel functions of second kind \cite{Gradshteyn}, then
\bea
\Im{\Pi_B(p)}&=&\frac{g^2p^2}{32\pi^2}\frac{4}{\sqrt{3}}\int^1_{\sqrt{1-\frac{4m^2}{p^2}}} dv v^2 \frac{\sqrt{m^2-\frac{1}{4}(1-v^2)p^2}}{(1-v^2)|p_\perp|eB} K_{1/3}(\rho).
\eea
By using the asymptotic behavior of the Modified Bessel functions of second kind in the regions:
\\

\noindent$\rho\gg1$ (which corresponds to $\lambda \ll 1$)
\bea
     K_{1/3}(\rho)\approx\sqrt{\frac{\pi}{2}} \rho^{-1/2} e^{-\rho},
\eea
\noindent$\rho\ll1$ (which corresponds to $\lambda \gg 1$)
\bea
     K_{1/3}(\rho)\approx\frac{\Gamma\left(\frac{1}{3}\right)}{2^{2/3}}\rho^{-1/3},
\eea
the imaginary part has respectively the form:
\bea
     \Im{\Pi_B(p)}&=&\frac{g^2p^2}{32\pi}\frac{\sqrt{3}}{m^2}\frac{1}{\sqrt{2\pi \lambda}}\int^1_{\sqrt{1-\frac{4m^2}{p^2}}} dv  \frac{v^2}{\sqrt{1-v^2}}\frac{1}{\left(1-(1-v^2)\frac{p^2}{4m^2}\right)^{1/4}} \exp\left[{-\frac{4}{\lambda}\frac{\left(1-(1-v^2)\frac{p^2}{4m^2}\right)^{3/2}}{1-v^2}}\right]
\label{lambdasmall}
\eea
and 
\bea
     \Im{\Pi_B(p)}&=&\frac{g^2p^2}{32\pi^2}\frac{\sqrt{3}\Gamma(\frac{1}{3})}{2^{1/3}}\frac{1}{m^2\lambda^{2/3}}\int^1_{\sqrt{1-\frac{4m^2}{p^2}}} dv  \frac{v^2}{(1-v^2)^{2/3}}.
\label{lambdalarge}
\eea

Since the integration over $v$ in Eq.(\ref{lambdasmall}) cannot be performed analytically,  we solved it numerically and plotted it in Fig.(\ref{figalatsai})(a).  On the other hand,  the integration over $v$ in Eq.(\ref{lambdalarge}) is easily done analytically and its behavior is shown in Fig.(\ref{figalatsai})(b).
\begin{figure}[h!]
\begin{tabular}{cc}
  \includegraphics[width=0.48\textwidth]{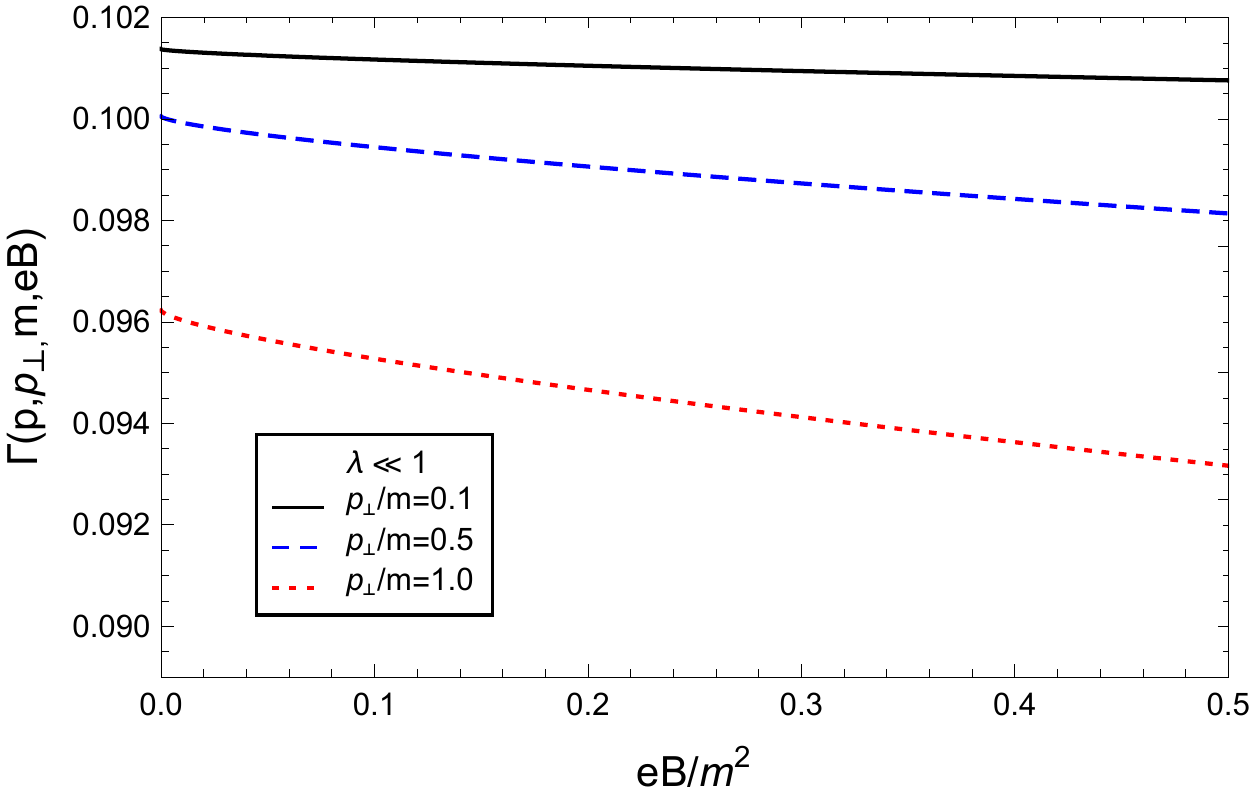}  &
  \includegraphics[width=0.48\textwidth]{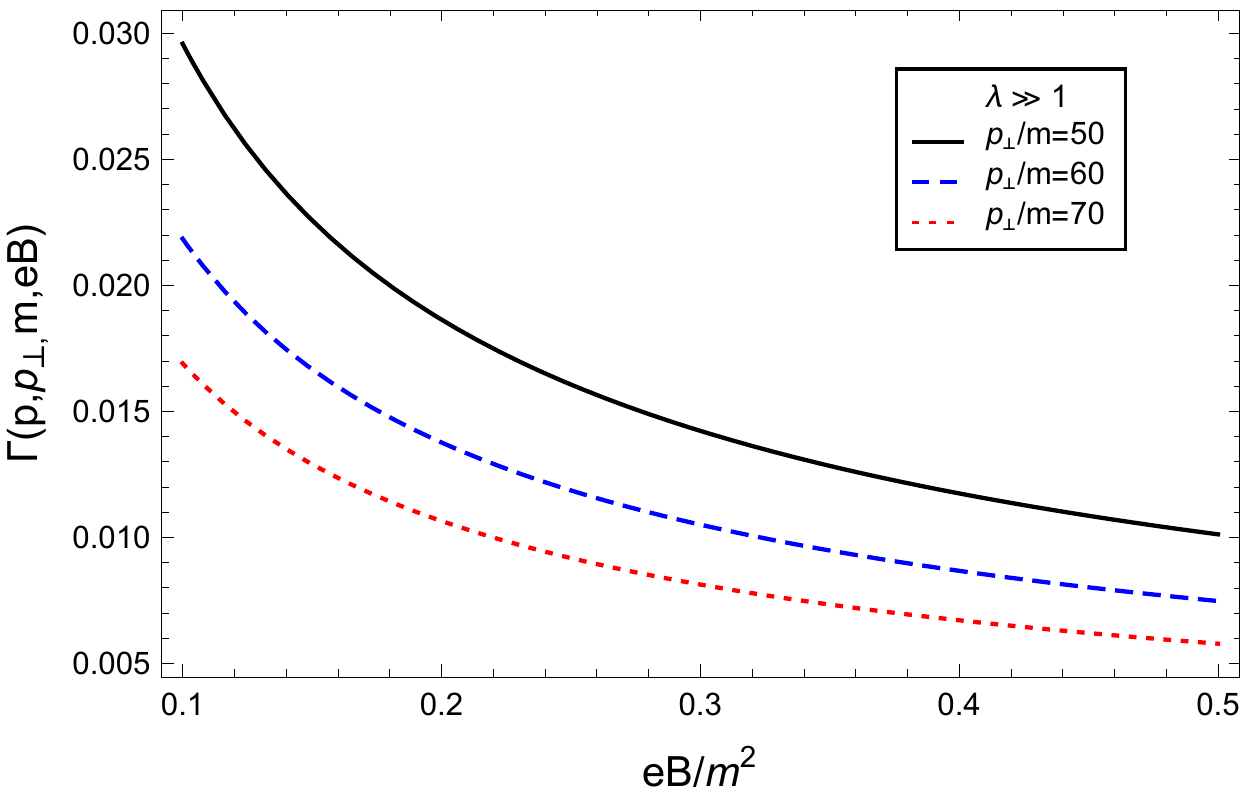} \\
  (a) & (b) \\
\end{tabular}
\caption{(color online) Decay width as a function of $eB/m^2$ in the regions (a) $\lambda \ll1$ and (b) $\lambda \gg 1$, both for several values of $p_\perp$.}
\label{figalatsai}
\end{figure}
The limit $eB\rightarrow0$ of Eq.(\ref{lambdasmall}) is shown in Appendix \ref{diracdeltalimit}.

At this point let us remark the differences between the results we have found in this work and the results reported by other authors in literature. On one hand, the self-energy that allows us to study the process $\Phi\rightarrow\phi^-+\phi^+$, by mean of the optical theorem, does not account for any spin, meanwhile, the photon polarization operator (PPO) and neutrino self-energy, that contain information about the processes $\gamma \rightarrow e^-+e^+$ and $\nu\rightarrow e^-+W^+$, respectively, take into account both internal and external particle spins. This observation suggests that particles spin play an important role in the analytical structure of the self-energy. An intuitive analysis supporting this idea could be based on the ultraviolet divergences:  both  PPO and neutrino self-energy depend on momenta as $1/pq$ and $1/pq^2$ exhibiting quadratic and  linear divergences, respectively, while in the present work, the self-energy divergences logarithmically due to $1/p^2q^2$. Thus, as the magnetic field is a physical scale, then it is expected to appear in different ways throughout the different cases studied in the literature.

On the other hand, note that in this approximation, when the transverse momentum of the decaying particle goes to zero, the effect of the magnetic field disappears. This could be due to the fact that at the beginning it is assumed that $\lambda \propto p_\perp eB$ can be large enough to prevent a power $eB$ expansion of the exponential, however this analysis breaks down when $\lambda\ll 1$ \cite{FelixKarbstein}. In this region, in order to be consistent, we should expand all factors that involve the magnetic field in powers of $eB$ up to $(eB)^2$. 

\subsection{Weak field limit as insertion of two photons}

Let us analyze the decaying particle with small transverse momentum in the presence of a weak external magnetic field. Starting with Eq.(\ref{selfenergyfull}), we expand all terms up to $(eB)^2$, as done in Sec.\ref{sec2}, getting
\bea
    \Pi_B(p)=i\frac{g^2 p^2}{32\pi^2}\int_{0}^1dv v^2 \int_0^\infty ds\left[1+\frac{1}{6}(eBs)^2\left(1-(1-v^2)\frac{p_\perp^2}{p^2}\right)-\frac{i}{48}s(eBs)^2(1-v^2)^2p_\perp^2\right]e^{is[\frac{1}{4}(1-v^2)p^2-m^2]},
 \nonumber \\
\eea
where we dropped the ultraviolet divergent part.

Once we perform the integration over  $v$ and the proper time $s$, the imaginary part of the self-energy reads
\bea
     &&\hspace{-0.5cm}\Im{\Pi_B(p)}=\frac{g^2}{16\pi}\sqrt{1-\frac{4m^2}{p^2}}
     \nonumber \\
     &&\times \left\{1-\frac{2}{3}\left(\frac{eB}{m^2}\right)^2\left(\frac{m^2}{p^2}\right)^2\frac{1}{\left(1-\frac{4m^2}{p^2}\right)^2}\left[1-2\left(\frac{p_\perp}{m}\right)^2\left(\frac{m^2}{p^2}\right)\frac{3+8\frac{m^2}{p^2}-14\frac{m^4}{p^4}}{\left(1-\frac{4m^2}{p^2}\right)}\right]\right\}\theta(p^2-4m^2).
\eea
In the above equation it is clear that the magnetic field modifies the imaginary part of the self-energy even when the decaying particle is at rest, suppressing it. On the other hand, the transverse momentum acts in the opposite direction, enhancing it. This behavior is shown in Fig.\ref{Alanosotros}: (a) Shows the imaginary part of the self-energy and (b) the decay width, both as a function of $eB/m^2$ for three different transverse momentum values:  $p_\perp/m=0,\ 0.5\ \mbox{and}\ 1$. Notice that in figure (a) the effect of the transverse momentum is highlighted, whereas in figure (b) the time dilatation by the Lorentz suppression factor is emphasized. This behavior is expected because a particle moving in the presence of a uniform magnetic field is also under the influence of an external electric field in its rest frame.
\begin{figure}[h!]
\begin{center}
\begin{tabular}{cc}
  \includegraphics[width=0.48\textwidth]{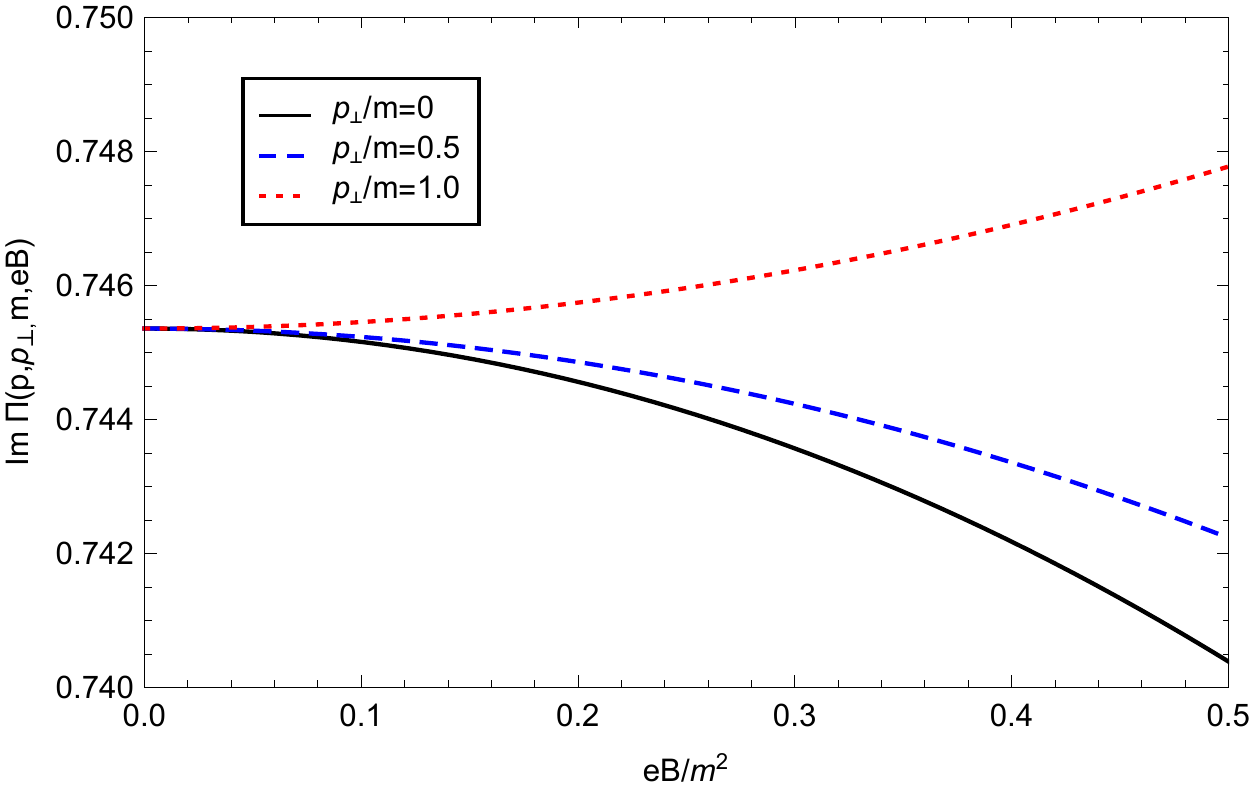}  &
  \includegraphics[width=0.48\textwidth]{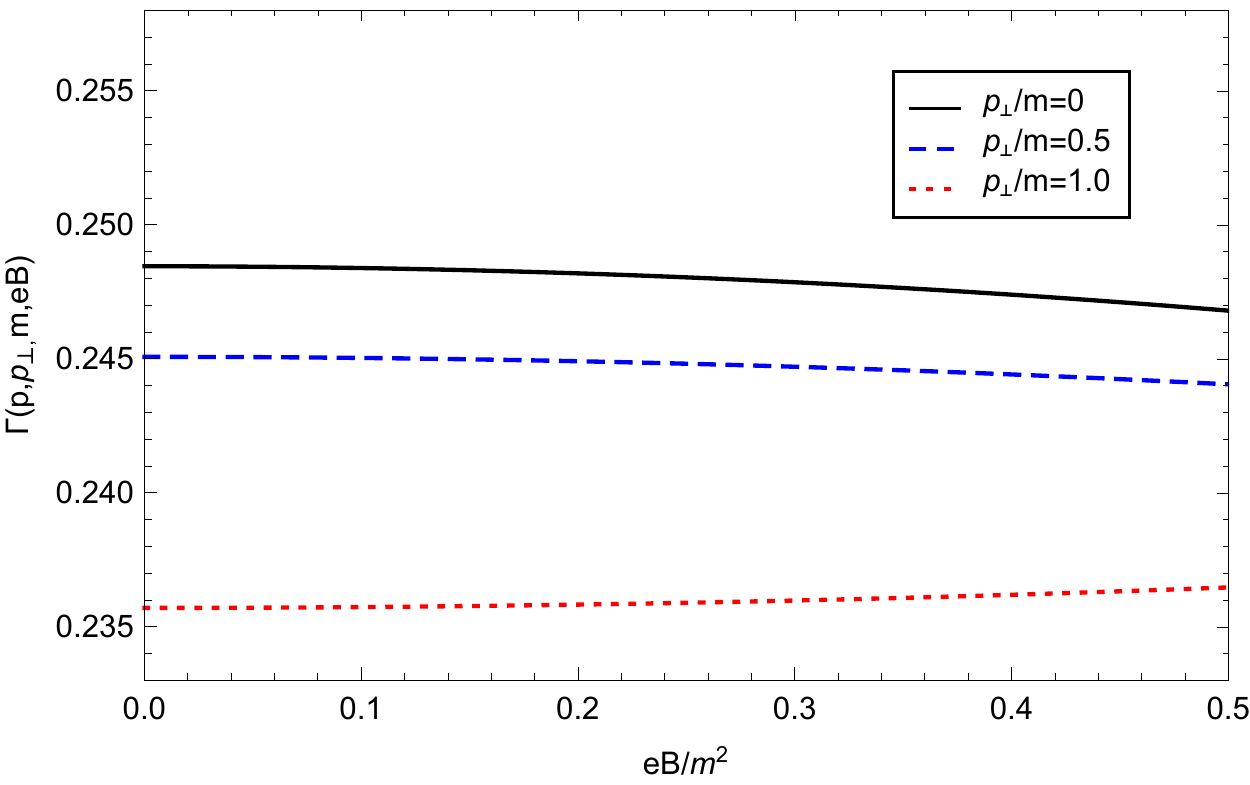} \\
  (a) & (b)
\end{tabular}
    \end{center}
    \caption{(color online) (a) Shows the imaginary part of the self-energy  and (b) the decay width, both as a function of $eB/m^2$ for three different transverse momentum values:  $p_\perp/m=0$ (continuous line), $p_\perp/m=0.5$ (dashed line) and $p_\perp/m=1$ (dotted line).}
\label{Alanosotros}
\end{figure}

\section{Conclusions}\label{sec5}

In this work we have studied the magnetic field and thermal effects on the decay process  of a neutral scalar boson into two charged scalar bosons. Focusing on a weak magnetic field, we found that as the magnetic field strength increases the pair creation is inhibited, meanwhile temperature has the opposite effect.

Since in this calculations some kind of approximation is needed, we employed a perturbative approach to obtain our main result. In order to explore whether the behavior we found could be associated with the employed approximation, we went through the same calculation with another approximation method, extensively used in  literature, for an specific case: scalar decay in vacuum. We still found that scalar decay width is inhibited by the magnetic field.

Although the suppression due to magnetic field on the scalar decay is present in the two employed methods, its behavior is different with respect to the magnetic field strength: in one case it is more pronounced. From the physical point of view, this difference could be due to the fact that the crossed approximation can be related with an infinite photon insertion meanwhile the perturbative approximation considers only  two photons, in analogy with Furry's theorem. 

Following the crossed field approximation, the results that can be found in the literature for the processes $\nu\rightarrow W^+ +\ e^-$~\cite{Erdas,Bhattacharya,Kuznetsov1} and $\gamma\rightarrow e^+ +\ e^-$~\cite{TsaiErber,Urrutia} show that the magnetic field enhances the pair creation, which is the opposite behavior of our findings. The difference can be traced back to the analytical structure of the self-energy imposed by the spin particles involved in the processes. An intuitive analysis supporting this idea goes as follows: in these two processes, the divergences are linear and quadratic, respectively,  meanwhile in the case studied in this work, $\Phi\rightarrow \phi^+ +\ \phi^-$, the self-energy diverges logarithmically. These different dependences on the momentum in the loop at the end will determine the response to the external magnetic field. The role played by the spin in the pair creation process in a magnetic field background is a work in progress and will be reported elsewhere.

\appendix

\section{Scalar decay in vacuum}

In this appendix, we calculate the imaginary part of the scalar self-energy in vacuum for equal masses $m_1=m_2=m$, by using standard Feynman rules 
\begin{equation}
i\Pi(p)=(ig)^2\int \frac{d^4k}{(2\pi)^4}D(k)D(p-k),
\end{equation}
where 
\begin{equation}
    D(k)=\int_0^\infty dse^{is(k^2-m^2+i\epsilon)}
\end{equation}
is the scalar propagator written in terms of Schwinger proper time.

In this way the scalar self-energy has the form
\begin{equation}
i\Pi(p)=-g^2\int \frac{d^4k}{(2\pi)^4}\int_0^\infty ds_1 ds_2
          e^{-i(s_1+s_2)m^2} 
          e^{is_1(p-k)^2}
          e^{is_2k^2}.
\end{equation}
Once we perform the gaussian integration over the loop momentum $k$, we get
\bea
\Pi(p)=\frac{g^2}{16\pi^2}\int_0^\infty ds_1 ds_2 \frac{1}{(s_1+s_2)^2}
          e^{-i(s_1+s_2)m^2} 
          e^{i \frac{s_1s_2}{s_1+s_2}p^2}.
\eea
Making use of the variables
\bea
      s_1=s\frac{1-v}{2}  \hspace{1.5cm}\mbox{and}\hspace{1.5cm}
      s_2=s\frac{1+v}{2},
\eea
the scalar self-energy can be rewritten as
\bea
     \Pi(p)=\frac{g^2}{16\pi^2}\int_{0}^1dv 
          \int_0^\infty \frac{ds}{s}e^{is \left(\frac{1}{4}(1-v^2)p^2-m^2\right)}.
\eea
The imaginary part of this expression reads
\bea
   \Im\Pi(p)=\frac{g^2}{16\pi}\int_{0}^1dv 
          \frac{1}{2\pi i}\int_{-\infty}^\infty \frac{ds}{s}e^{is \left(\frac{1}{4}(1-v^2)p^2-m^2\right)}.
\eea
Identifying the integration over $s$ as the integral representation of the Heaviside step function, we finally arrive at 
\bea
     \Im\Pi(p)&=&\frac{g^2}{16\pi}\int_{0}^1dv 
         \ \theta\left[\frac{1}{4}(1-v^2)p^2-m^2\right]
\nonumber \\
           &=&\frac{g^2}{16\pi}\sqrt{1-\frac{4m^2}{p^2}}\theta(p^2-4m^2) .
\label{ImPivacuum}    
\eea

\section{$eB\rightarrow0$ limit of Eq.(\ref{lambdasmall})} \label{diracdeltalimit}

In this appendix, we take the  zero magnetic field limit $eB\rightarrow 0$ in the case $\lambda \ll 1$, in order to compare it with Eq.(\ref{ImPivacuum}).

Starting with Eq.(\ref{lambdasmall}) 
\bea
 \Im{\Sigma(p)}&=&\frac{g^2p^2}{2(4\pi)^2}\frac{\sqrt{3}\pi}{m^2}\int^1_{\sqrt{1-\frac{4m^2}{p^2}}} dv  \frac{v^2}{\sqrt{1-v^2}}\frac{1}{\left(1-(1-v^2)\frac{p^2}{4m^2}\right)^{1/4}} \frac{1}{\sqrt{2\pi \lambda}}\exp\left[{-\frac{4}{\lambda}\frac{\left(1-(1-v^2)\frac{p^2}{4m^2}\right)^{3/2}}{1-v^2}}\right]
\label{imlimitdelta}
\eea
and noticing that in the limit $\lambda\rightarrow 0$ ($eB \rightarrow 0$), the last two factors in the integrand form a representation of the Dirac delta function, that is
\bea
    \delta(x)=\lim_{\epsilon\rightarrow0^+}\frac{1}{\sqrt{2\pi \epsilon}}\exp\left(-\frac{x^2}{2\epsilon}\right),
\eea
then Eq.(\ref{imlimitdelta}) can be written as
\bea
\Im{\Sigma(p)}&=&\frac{g^2p^2}{2(4\pi)^2}\frac{\sqrt{3}\pi}{m^2}\int^1_{\sqrt{1-\frac{4m^2}{p^2}}} dv  \frac{v^2}{\sqrt{1-v^2}}\frac{1}{\left(1-(1-v^2)\frac{p^2}{4m^2}\right)^{1/4}} \delta\left[{2\sqrt{2}\frac{\left(1-(1-v^2)\frac{p^2}{4m^2}\right)^{3/4}}{\sqrt{1-v^2}}}\right].
\eea
By using the Dirac delta function properties 
\bea
      \delta[\alpha x(v)]=\frac{1}{|\alpha|}\delta[x(v)],  \hspace{1cm}  \delta[x(v)]=\frac{\delta(v-v_0)}{\left.|x'(v)\right|_{v=v_0}},
\eea
where $v_0=\sqrt{1-\frac{4m^2}{p^2}}$, we obtain
\bea
\Im{\Sigma(p)}&=&\frac{g^2}{16\pi}\sqrt{\frac{2}{3}}\sqrt{1-\frac{4m^2}{p^2}}.
\eea
This equation reproduces the vacuum result shown in Eq.(\ref{ImPivacuum}) up to a $\sqrt{2/3}$ factor. The difference could come from the weak field expansion that preserves several powers in $(eB)$ in the argument of the exponential in Eq.(\ref{tsai1}).

\acknowledgements
Support for this work has been received in part from
DGAPA-UNAM under grant numbers PAPIIT-IN117817 and PAPIIT-IA107017.


\end{document}